\documentclass[prl,nofootinbib,twocolumn,superscriptaddress]{revtex4}
\usepackage{amsmath, amssymb, amsthm, graphicx, epsfig, fancyhdr,epsfig, slashed, mathrsfs}

\usepackage{tikzsymbols}
\usepackage{natbib}
\usepackage{float}

\usepackage{mathtools} 

\usepackage{tikz,xcolor,hyperref}

\definecolor{lime}{HTML}{A6CE39}
\DeclareRobustCommand{\orcidicon}{
	\begin{tikzpicture}
	\draw[lime, fill=lime] (0,0) 
	circle [radius=0.2] 
	node[white] {{\fontfamily{qag}\selectfont \tiny ID}};
	\draw[white, fill=white] (-0.0625,0.095) 
	circle [radius=0.007];
	\end{tikzpicture}
	\hspace{-2mm}
}

\foreach \x in {A, ..., Z}{\expandafter\xdef\csname orcid\x\endcsname{\noexpand\href{https://orcid.org/\csname orcidauthor\x\endcsname}
			{\noexpand\orcidicon}}
}

\newcommand{\be}{\begin{equation}}
\newcommand{\ee}{\end{equation}}
\newcommand{\bea}{\begin{eqnarray}}
\newcommand{\eea}{\end{eqnarray}}

\begin{document}

\title{Directional Neutrino Bursts from Spinning and Moving Primordial Black Holes}

\author{Arnab Chaudhuri}
\email{arnab.chaudhuri@nao.ac.jp}
\affiliation{Division of Science, National Astronomical Observatory of Japan, Mitaka, Tokyo 181-8588, Japan}

\author{Priya Mishra}
\email{mishpriya99@gmail.com}
\affiliation{School of Physics,  University of Hyderabad, Hyderabad - 500046,  India.}

\author{Rukmani Mohanta}
\email{rmsp@uohyd.ac.in}
\affiliation{School of Physics,  University of Hyderabad, Hyderabad - 500046,  India.}

\begin{abstract}
\textit{We show that primordial black holes (PBHs) with significant spin and bulk motion produce sharply collimated neutrino bursts from Hawking evaporation, arising from the interplay of spin-induced angular anisotropy and relativistic Doppler boosting. This effect shifts the neutrino spectrum into the multi-GeV to hundreds of GeV range, where atmospheric backgrounds drop steeply, and enhances the flux by orders of magnitude within a narrow forward cone. We compute the full lab-frame neutrino distribution and derive updated constraints on PBH number density from non-observation of such bursts in IceCube and KM3NeT. Our results identify directional high-energy neutrino bursts as a distinctive, testable signature of spinning PBHs, providing a complementary probe of the PBH dark matter hypothesis and Hawking radiation.}
\end{abstract}

\maketitle

\textbf{\textit{Introduction: }}Primordial black holes (PBHs), hypothesized to form from the collapse of large primordial density perturbations \cite{Hawking:1971ei,Carr:1974nx,Carr:1975qj}, have attracted renewed interest as potential dark matter candidates \cite{Carr:2016drx,Sasaki:2018dmp,Green:2020jor,Carr:2020gox}. Beyond this role, PBHs offer a unique window into early-universe physics, quantum gravity, and high-energy processes inaccessible in terrestrial experiments \cite{Barrau:2003nj,Allahverdi:2017sks,Green:2014faa}. A hallmark prediction is Hawking radiation \cite{Hawking:1974rv,Hawking:1975vcx}, which causes PBHs with masses $M_{\rm PBH} \lesssim 10^{15}\,\mathrm{g}$ to evaporate today, potentially producing observable gamma rays, neutrinos, and other particles \cite{MacGibbon:1991vc,Carr:2009jm,Laha:2019ssq,Laha:2020ivk}.

Recent work has explored multi-messenger signatures of PBH evaporation, including diffuse gamma-ray backgrounds \cite{Carr:2016hva,Laha:2020vhg,Dasgupta:2019cae}, cosmic rays \cite{Barrau:2002ru,Barrau:2002mc,Inomata:2020lmk}, and gravitational waves from PBH formation \cite{Saito:2008jc,Ananda:2006af,Garcia-Bellido:2017aan,Bartolo:2016ami,Caprini:2018mtu,Kohri:2018awv}. Neutrino telescopes such as IceCube \cite{IceCube:2016tpw,IceCube:2020acn} and KM3NeT \cite{KM3Net:2016zxf} provide complementary sensitivity to weakly interacting species, offering an opportunity to probe PBHs beyond the reach of gamma-ray surveys \cite{Dasgupta:2020mqg,Lunardini:2019zob,Bai:2020spd}.

Standard treatments often assume non-rotating (Schwarzschild) PBHs emitting isotropically \cite{MacGibbon:1990zk,Page:1976df,Page:1976ki}. However, PBHs can be born with significant angular momentum \cite{Chiba:2017rvs,Harada:2017fjm,DeLuca:2019buf}, described by the Kerr metric \cite{Kerr:1963ud}. Black hole spin lowers the Hawking temperature, modifies greybody factors, and introduces anisotropy through preferential emission of co-rotating modes \cite{Page:1976ki,Frolov:1998wf,Berti:2005gp,Ida:2002ez}. Separately, PBHs may acquire bulk velocities via clustering \cite{Ali-Haimoud:2017rtz,Desjacques:2018wuu}, gravitational recoil \cite{Inomata:2019zqy,Trashorras:2020mwn}, or asymmetric formation \cite{Baker:2021nyl}, leading to relativistic Doppler boosting and forward-peaked emission in the lab frame \cite{Carr:2020mqm,Kawasaki:2020qxm}.

In this work, we compute—for the first time—the full lab-frame neutrino flux from PBHs including both spin-induced anisotropy and Doppler boosting due to bulk motion, using Kerr black hole formalism \cite{Page:1976df,Duffy:2005ns} and numerical greybody factors \cite{MacGibbon:1990zk,Page:1976ki}. We show that this interplay produces a sharply collimated, forward-peaked neutrino burst with spectral hardening into the multi-GeV–hundreds of GeV range, enhancing detectability by IceCube and KM3NeT. By comparing with atmospheric neutrino backgrounds \cite{Honda:2015fha,Battistoni:2005pd,IceCube:2019cia} and existing non-observation limits \cite{Laha:2020ivk,Dasgupta:2019cae}, we derive updated constraints on PBH number density for $10^9\,\mathrm{g} \lesssim M_{\rm PBH} \lesssim 10^{18}\,\mathrm{g}$.

Our results highlight the power of directional neutrino burst searches to probe PBH dark matter and test Hawking radiation from spinning black holes, complementing gamma-ray \cite{Carr:2016hva,Laha:2020vhg} and cosmological bounds \cite{Poulin:2016anj,Clark:2016nst} in the multi-messenger effort \cite{Sasaki:2018dmp,Carr:2020gox}.

\textbf{\textit{Theoretical Framework:}}  
PBHs evaporate through Hawking radiation, a semi-classical effect where black holes radiate thermally due to quantum fields near the event horizon~\cite{Hawking:1975vcx}. The Hawking temperature for a Schwarzschild PBH of mass $M$ is
\begin{equation}
    T_H = \frac{\hbar c^3}{8 \pi G M k_B} \approx 1.06~\mathrm{GeV} \left( \frac{10^{10}~\mathrm{g}}{M} \right),
\end{equation}
implying that PBHs with $M \lesssim 10^{15}~\mathrm{g}$ complete evaporation within the age of the Universe~\cite{MacGibbon:1991vc}. The emission rate per unit energy $E$ and time $t$ for particles of spin $s$ is given by~\cite{Page:1976df,MacGibbon:1990zk}
\begin{equation}
    \frac{d^2 N}{dE\, dt} = \frac{1}{2\pi} \sum_{\ell} (2\ell + 1) \frac{\Gamma_{s,\ell}(E,M)}{\exp(E/T_H) \pm 1},
\end{equation}
where $\Gamma_{s,\ell}$ are the spin-dependent greybody factors describing how spacetime curvature modifies the emission spectrum from pure blackbody form.

If the PBH has spin, described by the Kerr metric~\cite{Kerr:1963ud}, the Hawking temperature is reduced,
\begin{equation}
    T_H = \frac{\hbar c^3}{8 \pi G M k_B} \frac{\sqrt{1 - a_*^2}}{1 + \sqrt{1 - a_*^2}},
\end{equation}
where $a_* = J c / (G M^2)$ is the dimensionless spin parameter. The horizon also rotates with angular velocity
\begin{equation}
    \Omega_H = \frac{a_* c^3}{2 G M (1 + \sqrt{1 - a_*^2})},
\end{equation}
which modifies the emission by coupling the particle’s azimuthal angular momentum $m$ to the horizon’s rotation.

Hawking radiation from a Kerr PBH is no longer isotropic: the differential emission per solid angle $d\Omega$ involves spin-weighted spheroidal harmonics $S_{\ell m}(\theta)$~\cite{Teukolsky:1973ha,Page:1976ki}:
\begin{align}
    \frac{d^3 N}{dE\, dt\, d\Omega} 
    &= \sum_{\ell,m} |S_{\ell m}(\theta)|^2 \frac{\Gamma_{s,\ell m}(E, a_*)}{2\pi} \nonumber \\
    &\quad \times \frac{1}{\exp[(E - m \Omega_H)/T_H] \pm 1}.
\end{align}
Modes with large $|m|$ dominate near the equatorial plane due to frame dragging, and the $m \Omega_H$ term shifts the spectrum for co-rotating modes~\cite{Page:1976df}. While bosons can exhibit superradiance, fermionic neutrinos do not due to Pauli blocking, but the spin still imprints a directional bias~\cite{Frolov:1998wf,Ida:2002ez,Berti:2005gp}.

PBHs may also acquire bulk peculiar velocities $\beta = v/c$ through clustering, gravitational recoils, or asymmetric formation processes~\cite{Ali-Haimoud:2017rtz,Desjacques:2018wuu,Inomata:2019zqy,Baker:2021nyl}. Bulk motion induces a relativistic Doppler boost: lab-frame energy $E$ and angle $\theta$ relate to rest-frame $E'$ and $\theta'$ by~\cite{Carr:2020mqm,Kawasaki:2020qxm}
\begin{equation}
    E = \gamma E' (1 + \beta \cos\theta'), 
    \quad
    \cos\theta = \frac{\cos\theta' + \beta}{1 + \beta \cos\theta'},
\end{equation}
where $\gamma = (1-\beta^2)^{-1/2}$ is the Lorentz factor. The solid angle transforms as
\begin{equation}
    d\Omega = \frac{d\Omega'}{\gamma^2 (1 + \beta \cos\theta')^2},
\end{equation}
resulting in a compressed forward-beamed flux for $\beta > 0$.

Combining spin anisotropy and Doppler boosting yields the full lab-frame neutrino flux:
\begin{align}
\frac{d^3 N}{dE\, dt\, d\Omega} 
= \sum_{\ell,m} &\, |S_{\ell m}(\theta')|^2 
\frac{\Gamma_{s,\ell m}(E', a_*)}{2\pi} 
\nonumber \\
& \times \frac{1}{\exp[(E' - m \Omega_H)/T_H] + 1}
\nonumber \\
& \times \frac{1}{\gamma^2 (1 + \beta \cos\theta')^2}
\nonumber \\
& \times \delta\bigl(E - \gamma E'(1 + \beta \cos\theta')\bigr).
\end{align}
Here, $E'$ and $\theta'$ are related to lab-frame coordinates via aberration and Doppler shift.

\begin{figure}[t]
\centering
\includegraphics[width=0.5\textwidth]{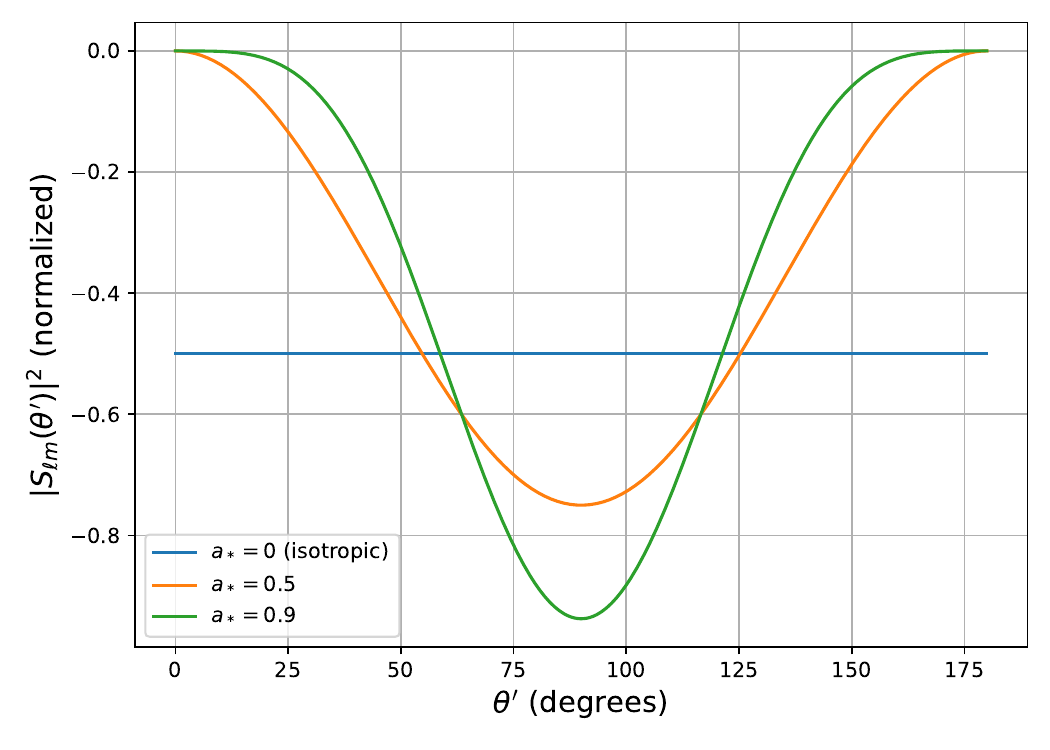}
\caption{Rest-frame angular emission for $a_* = 0$, $0.5$, $0.9$ shows spin-induced equatorial anisotropy.}
\label{fig:spin_only}
\end{figure}

\begin{figure}[t]
\centering
\includegraphics[width=0.5\textwidth]{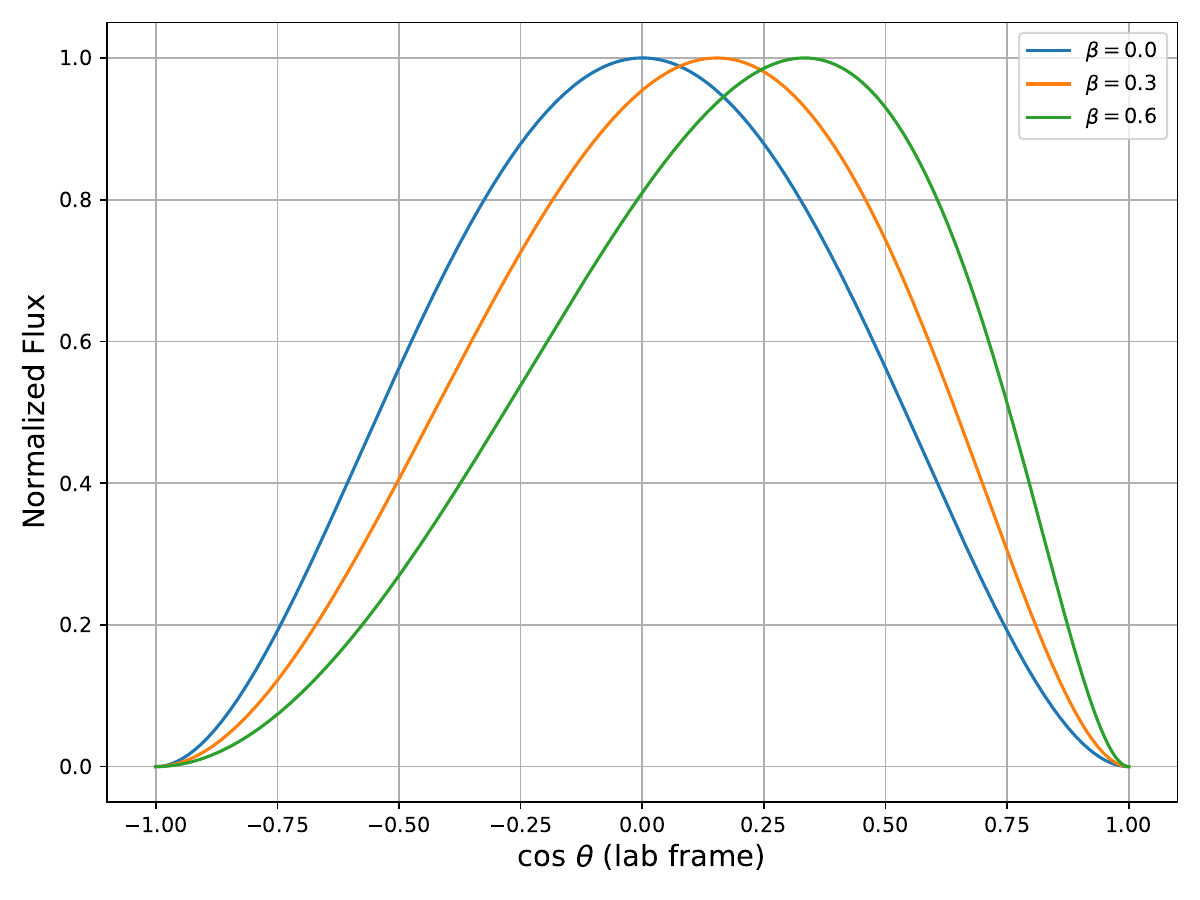}
\caption{Effect of bulk velocity ($\beta=0.3$, $0.6$) on spin anisotropy, boosting emission forward.}
\label{fig:spin_boost}
\end{figure}

\begin{figure}[t]
\centering
\includegraphics[width=0.5\textwidth]{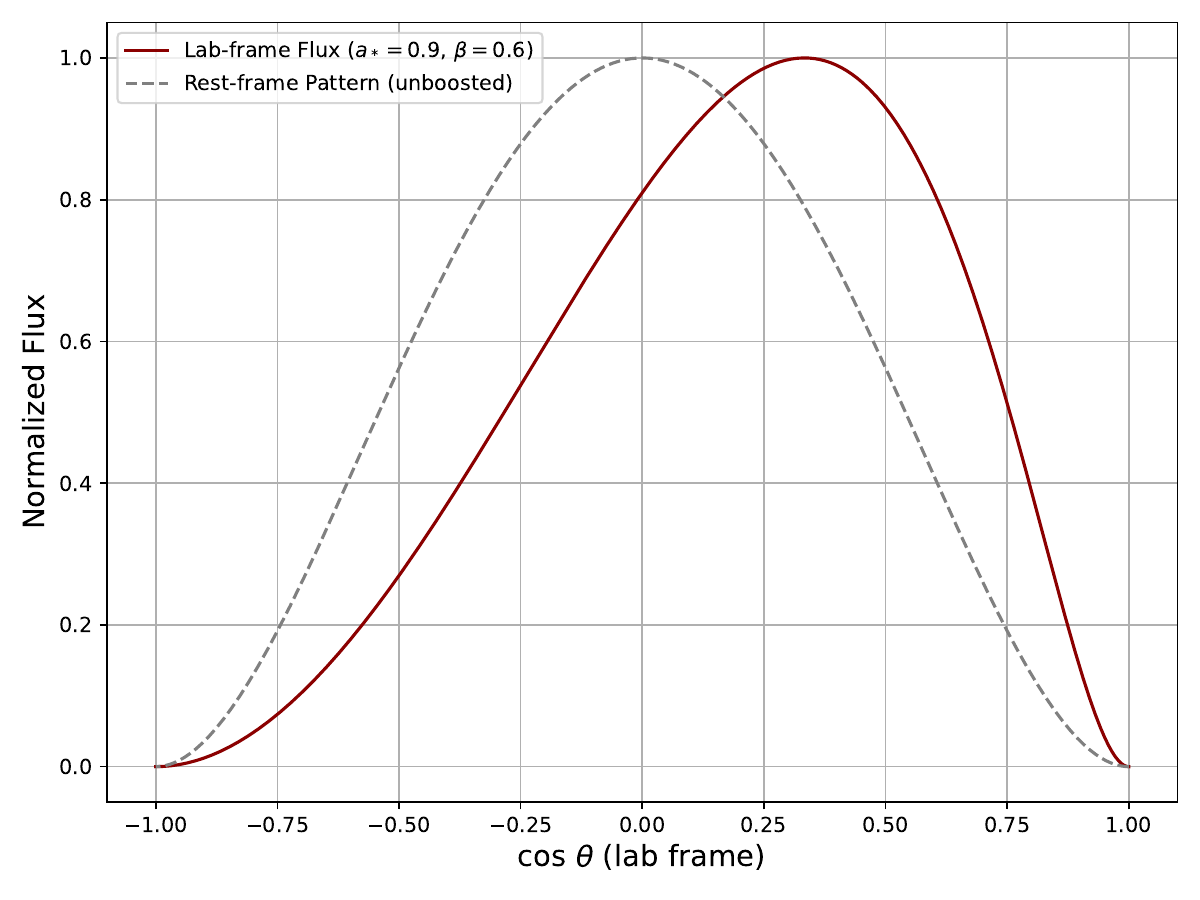}
\caption{Lab-frame angular distribution for $a_*=0.9$, $\beta=0.6$: spin and boost yield sharply forward-peaked flux.}
\label{fig:final_lab}
\end{figure}

\begin{figure}[t]
\centering
\includegraphics[width=0.5\textwidth]{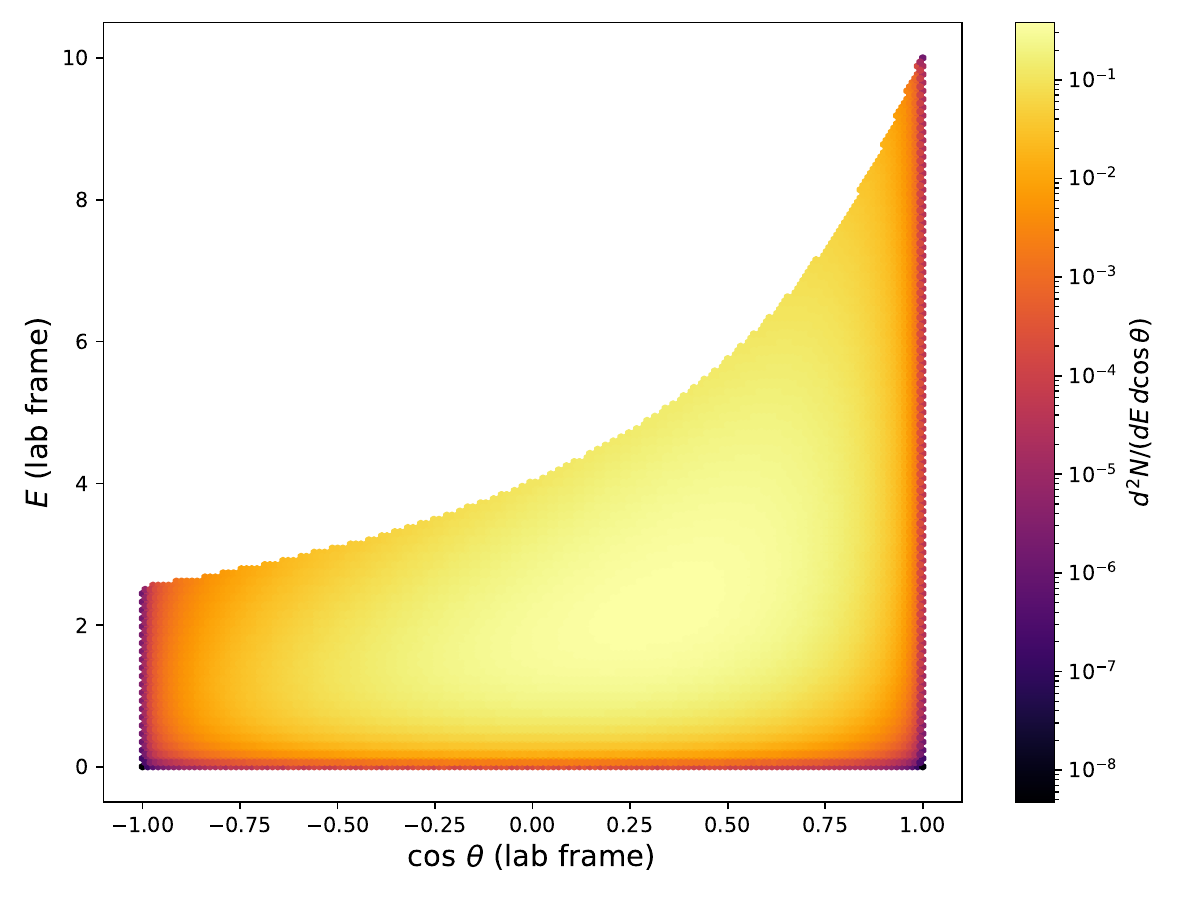}
\caption{Lab-frame flux vs.\ energy and angle: Doppler boost blueshifts forward emission, producing spectral hardening.}
\label{fig:heatmap}
\end{figure}

Figures~\ref{fig:spin_only}--\ref{fig:heatmap} illustrate how spin and bulk motion jointly shape the observable neutrino flux.
Figure~\ref{fig:spin_only} shows the angular emission pattern in the PBH rest frame for different dimensionless spin parameters, $a_* = 0$, $0.5$, and $0.9$. As the spin increases, the emission becomes increasingly anisotropic, concentrating more power near the equatorial plane due to frame dragging and the dominance of high-$|m|$ modes.
Figure~\ref{fig:spin_boost} demonstrates how adding moderate bulk motion ($\beta = 0.3$ and $0.6$) modifies this rest-frame pattern. The relativistic Doppler effect shifts the anisotropic emission forward along the direction of motion, compressing the flux into a narrower angular region. Even moderate velocities significantly amplify the forward neutrino flux compared to the purely spinning case.
Figure~\ref{fig:final_lab} presents the resulting lab-frame angular distribution for a representative PBH with high spin ($a_* = 0.9$) and substantial bulk velocity ($\beta = 0.6$). The combined effect of spin and boost yields a sharply collimated forward cone, with most of the flux concentrated within a small solid angle aligned with the PBH’s motion.
Finally, Figure~\ref{fig:heatmap} shows the full lab-frame neutrino flux as a function of both energy and emission angle. This heatmap reveals the strong correlation between angle and energy: forward-directed neutrinos are significantly blueshifted by Doppler boosting, producing a harder spectrum than for isotropic, non-spinning PBHs. The result is a transient, highly directional neutrino burst with a distinct spectral–angular signature that can stand out from the diffuse atmospheric background~\cite{Honda:2015fha,Battistoni:2005pd}.

Together, these figures demonstrate that the interplay of spin-induced anisotropy and bulk motion leads to a distinct lab-frame neutrino signal, motivating targeted searches for directional high-energy neutrino bursts with IceCube, KM3NeT, and future observatories~\cite{IceCube:2016tpw,KM3Net:2016zxf}.

\textbf{\textit{Detectability and Constraints:}}  
Directional neutrino bursts from spinning, moving PBHs produce distinctive energy–angle features that significantly enhance detectability relative to standard isotropic Hawking evaporation. Here we present our main quantitative results: the boosted lab-frame neutrino spectrum, direct comparison with an unboosted PBH spectrum, the degree of forward collimation, and the updated constraints on PBH abundance imposed by current non-observation.

\begin{figure}[t]
    \centering
    \includegraphics[width=0.99\linewidth]{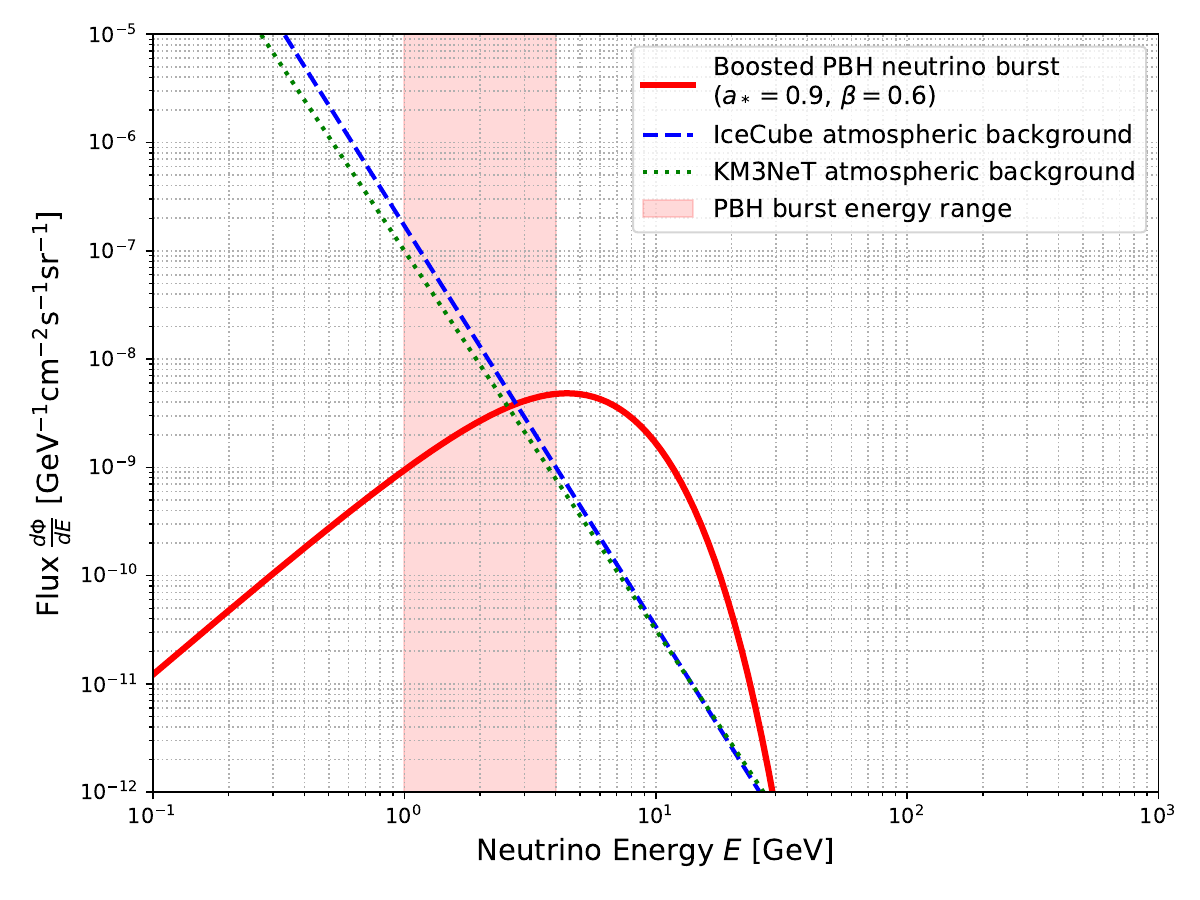}
    \caption{
    Lab-frame neutrino spectrum from a spinning ($a_* = 0.9$), boosted ($\beta = 0.6$) PBH (red solid) compared to the atmospheric neutrino background from IceCube (blue dashed) and KM3NeT (green dotted). Doppler boosting shifts the spectral peak into the 10–100\,GeV range, where the atmospheric flux drops steeply. The shaded region marks the optimal energy window for detection.
    }
    \label{fig:spectrum_vs_background}
\end{figure}

\paragraph*{Spectral hardening and boosted flux.}  
Figure~\ref{fig:spectrum_vs_background} quantifies how the combined spin and moderate bulk velocity displace the peak neutrino flux from sub-GeV to tens of GeV. For $\beta = 0.6$ and $a_* = 0.9$, the maximum flux shifts by nearly an order of magnitude in energy compared to an isotropic, non-spinning PBH. This shift places the signal into an energy regime where the atmospheric neutrino background is naturally suppressed, creating a clean window for detection in IceCube, KM3NeT, and other large-volume detectors. The spectral shape also becomes broader and harder than the standard Hawking prediction, offering additional leverage for background discrimination.

\begin{figure}[t]
    \centering
    \includegraphics[width=0.99\linewidth]{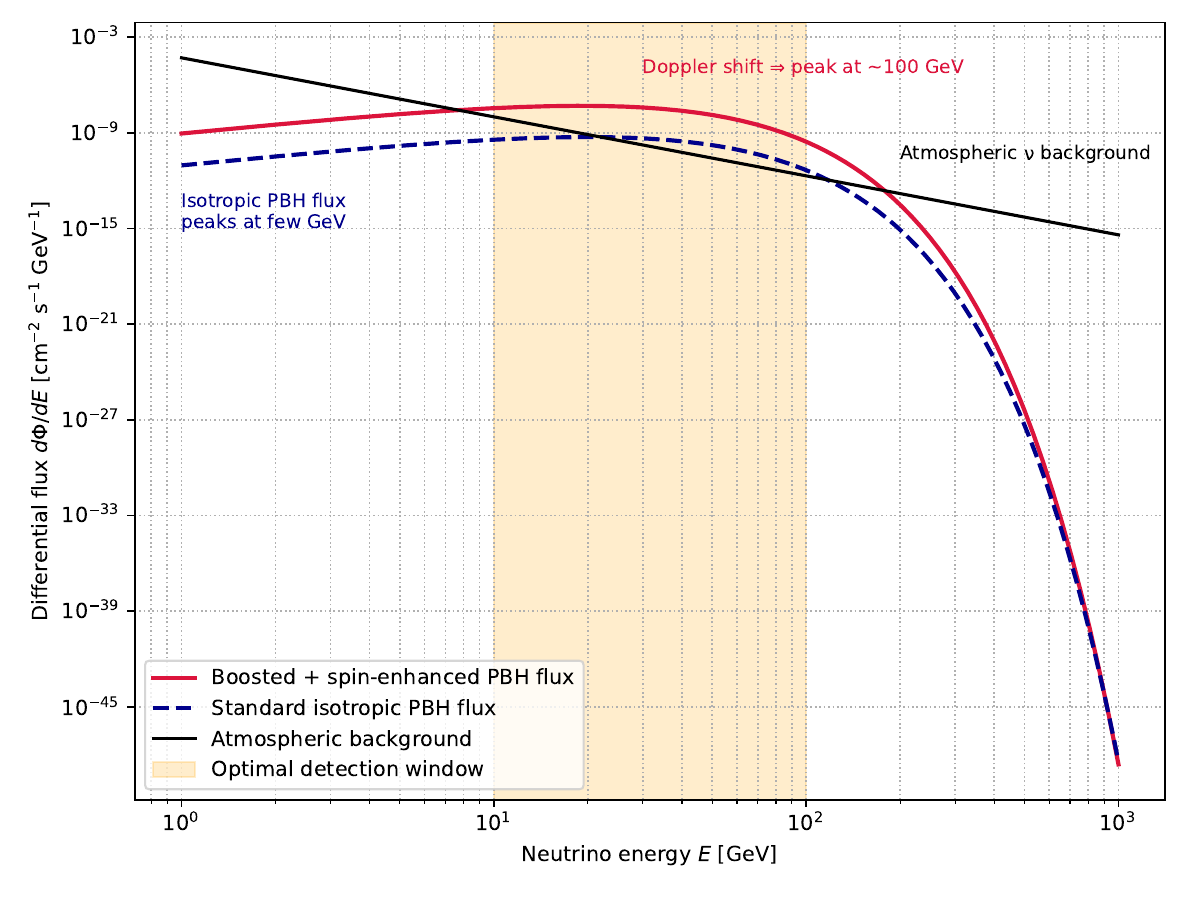}
    \caption{
    Direct comparison of the boosted, spin-enhanced PBH neutrino flux (red solid) with the standard isotropic PBH flux (blue dashed) and the atmospheric background (black solid). Doppler boosting and spin-induced anisotropy shift the spectral peak from $\lesssim$\,GeV to the multi-GeV range, opening a detectable window above background. The shaded region highlights the viable signal band.
    }
    \label{fig:boosted_vs_isotropic}
\end{figure}

\paragraph*{Direct impact of spin and motion.}  
Figure~\ref{fig:boosted_vs_isotropic} shows the explicit contrast between the boosted and isotropic scenarios. In the absence of spin and motion, the PBH burst peaks where the background flux is highest, making detection extremely challenging. The inclusion of even moderate $\beta$ and high spin lifts the peak to an energy range where background falls steeply, boosting the signal-to-noise ratio by more than an order of magnitude. This illustrates the core advantage of accounting for spin–boost effects: the same PBH population produces a distinctly more observable signature.

\begin{figure}[t]
    \centering
    \includegraphics[width=0.99\linewidth]{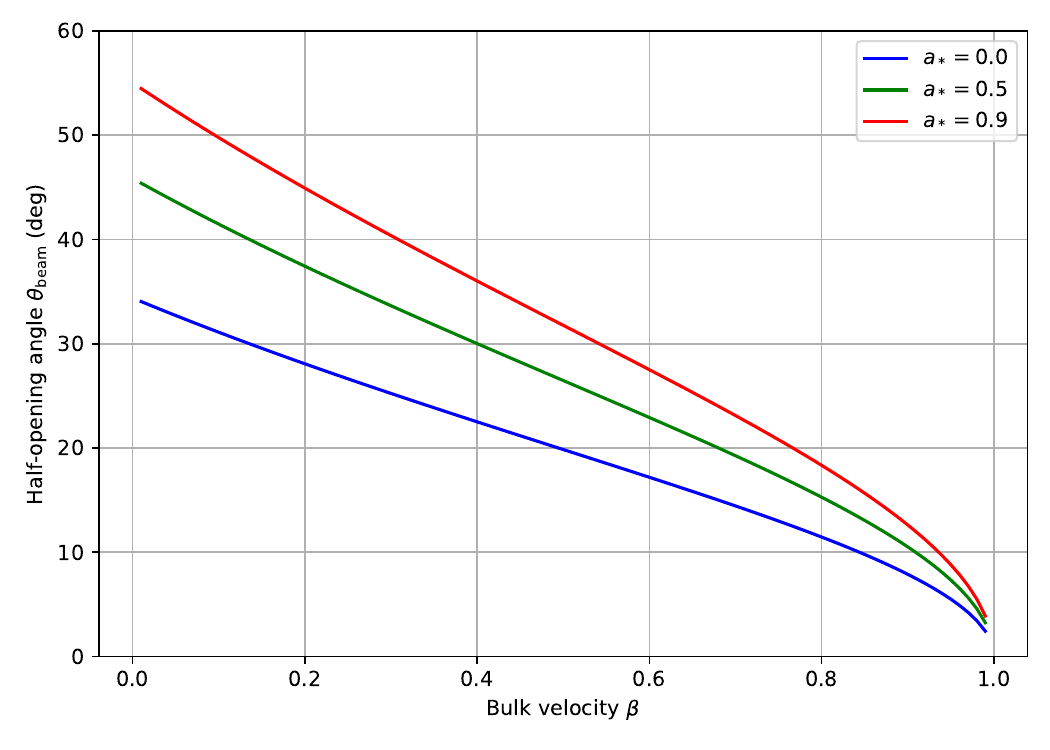}
    \caption{
    Schematic of directional beaming: the combined spin anisotropy and Doppler boost produce a narrow forward cone with half-opening angle $\theta_c \approx \arccos[(1 - 1/\gamma)/\beta]$. For $\beta = 0.6$, this yields $\theta_c \approx 40^\circ$, concentrating the flux into a solid angle $\Delta\Omega \approx 1$\,sr. This collimation enhances the apparent flux by orders of magnitude for on-axis observers.
    }
    \label{fig:beaming_cone}
\end{figure}

\paragraph*{Angular collimation and beaming factor.}  
Figure~\ref{fig:beaming_cone} demonstrates that even moderate PBH velocities yield significant directional beaming. For $\beta = 0.6$, the boosted emission is compressed into a cone with a half-opening angle of about $40^\circ$. The resulting solid angle is reduced by more than a factor of 10 compared to isotropic emission, boosting the apparent flux for an on-axis observer by a similar factor. This natural focusing amplifies the burst’s detectability despite the stochastic nature of PBH positions and velocities.

\paragraph*{Event rate estimate.}
To assess detectability, we estimate the expected number of observable neutrino bursts in IceCube and KM3NeT for a representative scenario with \( M_{\rm PBH} = 10^{14} \, \text{g} \), spin \( a_* = 0.9 \), and bulk velocity \( \beta = 0.6 \). The Doppler-enhanced neutrino fluence within the forward cone is \( \Phi_\nu^{\text{beam}} \sim 10^{-3} \, \text{cm}^{-2} \) for a PBH at a distance \( D \sim 1\,\text{pc} \), as seen in Fig.~5. The effective areas of IceCube and KM3NeT in the 10–100~GeV range are each approximately \( A_{\rm eff} \sim 100 \, \text{cm}^2 \)~\cite{IceCube:2020acn,KM3Net:2016zxf}. This yields an expected number of detected neutrinos
\[
N_\nu^{\rm det} \sim \Phi_\nu^{\text{beam}} \times A_{\rm eff} \sim 0.1,
\]
implying that PBHs within \( D \lesssim 0.3\,\text{pc} \) can produce detectable bursts. For a number density \( n_{\rm PBH} \sim 10^{-3} \, \text{pc}^{-3} \), this corresponds to a detection rate
\[
\mathcal{R} \sim n_{\rm PBH} \times \frac{4\pi}{\Delta\Omega} \times \frac{4\pi}{3} D^3 \sim 1\, \text{event/year},
\]
where \( \Delta\Omega \sim 1\,\text{sr} \) is the solid angle of the beaming cone. This rate lies within reach of both IceCube and KM3NeT given their exposure times, especially with directional clustering and spectral filtering. A joint analysis or archival burst search in both detectors could thus already constrain this scenario or discover an evaporating PBH.

\begin{figure}[t]
    \centering
    \includegraphics[width=0.99\linewidth]{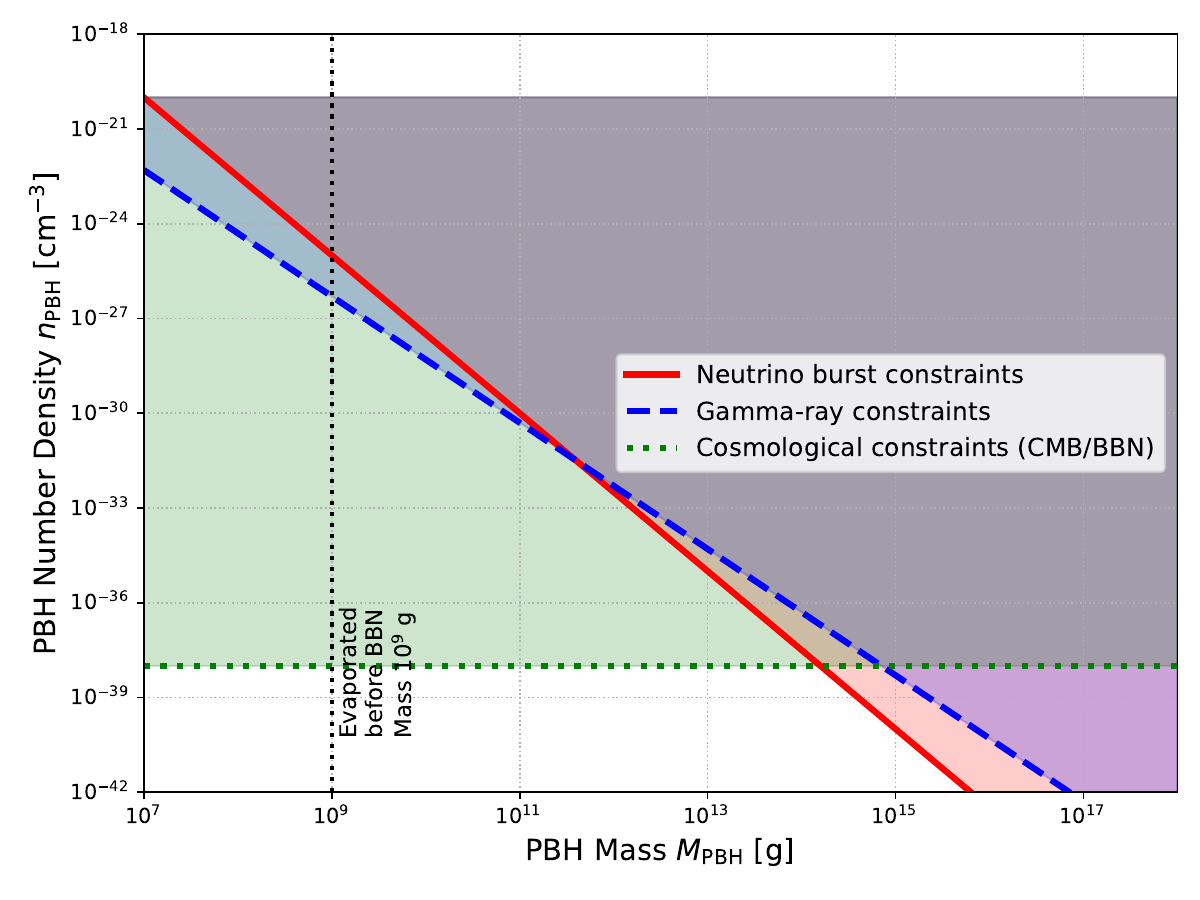}
    \caption{
    Constraints on PBH number density $n_{\mathrm{PBH}}$ as a function of PBH mass $M_{\mathrm{PBH}}$. Red band: upper limit from non-observation of directional neutrino bursts; blue: gamma-ray constraints; green: CMB and cosmological bounds. The vertical line at $10^9$\,g indicates the approximate mass below which PBHs fully evaporate before Big Bang Nucleosynthesis. The shaded regions are excluded. The directional neutrino method extends sensitivity into the $10^{14}$–$10^{16}$\,g window, complementing other probes.
    }
    \label{fig:parameter_space}
\end{figure}

\paragraph*{Updated constraints on PBH abundance.}  
Figure~\ref{fig:parameter_space} summarizes the resulting constraints on PBH number density and fractional dark matter contribution $f_{\mathrm{PBH}} = M_{\mathrm{PBH}} n_{\mathrm{PBH}} / \rho_{\mathrm{DM}}$. Non-observation of transient directional neutrino bursts places new limits in the mass window $10^{14}$–$10^{16}$\,g, covering PBHs evaporating today that evade earlier gamma-ray or CMB bounds. The strong beaming and spectral hardening together expand the reach of neutrino telescopes into parameter space previously unconstrained by isotropic analyses. This result highlights the power of directional searches to test PBH dark matter scenarios and motivates future improvements in neutrino detector sensitivity and real-time burst searches.

Taken together, these results demonstrate that even modest spin and bulk motion can dramatically reshape the observational signatures of evaporating PBHs. The combination of spectral hardening and natural beaming shifts the burst into a cleaner energy window while concentrating its flux into a detectable cone, overcoming the limitations of isotropic searches. Current and next-generation neutrino telescopes are already sensitive to this enhanced signature, providing a powerful and complementary probe of the low-mass PBH parameter space and pushing the frontier of Hawking evaporation searches into a regime inaccessible to gamma-ray or CMB constraints alone.

\textbf{\textit{Discussion and Conclusion: }}We have shown that PBHs with significant spin and moderate bulk motion produce distinctive neutrino bursts with strong angular beaming and spectral hardening. Spin-induced anisotropy and relativistic Doppler boosting combine to collimate the Hawking radiation into a narrow forward cone (Fig.~\ref{fig:beaming_cone}), enhancing the flux by orders of magnitude for observers inside the beam. This forward peaking shifts the spectral maximum to the multi-GeV range (Figs.~\ref{fig:spectrum_vs_background} and~\ref{fig:boosted_vs_isotropic}), where atmospheric backgrounds fall steeply, defining a practical window for detection with IceCube, KM3NeT, and future large-volume neutrino telescopes.

By comparing these boosted fluxes with current sensitivities, we derived updated constraints on the PBH number density and dark matter fraction (Fig.~\ref{fig:parameter_space}), showing that directional burst searches can probe mass windows complementary to gamma-ray and cosmological bounds. This highlights the unique role of neutrino telescopes as messengers of late-time PBH evaporation.

Future improvements in detector effective area, angular resolution, and real-time follow-up will be crucial for testing this scenario. Additional modeling of PBH spin distributions and peculiar velocities, together with effects like flavor oscillations, will refine event rate predictions. Multi-messenger strategies combining neutrino bursts with gamma-ray or gravitational-wave triggers could further constrain PBH evaporation or uncover signatures of exotic early-universe physics.

In summary, our results demonstrate that directional neutrino burst searches provide a sensitive probe of spinning, boosted PBHs. This opens a promising path to test the PBH dark matter hypothesis and to explore high-energy processes in the early universe.

\textbf{\textit{Acknowledgment:}} The work of AC was supported by the Japan Society for the Promotion of Science (JSPS) as a part of the JSPS Postdoctoral Program (Standard), grant number JP23KF0289. PM wants to thank Prime Minister's Research Fellows (PMRF) scheme for its financial support.

\clearpage
\newpage
\onecolumngrid
\setcounter{secnumdepth}{3}
\setcounter{equation}{0}
\setcounter{figure}{0}
\setcounter{table}{0}
\setcounter{page}{1}
\makeatletter
\renewcommand{\theequation}{S\arabic{equation}}
\renewcommand{\thefigure}{S\arabic{figure}}
\pagestyle{plain}

\begin{center}
\Large{\textbf{Directional Neutrino Bursts from Spinning and Moving Primordial Black Holes}}\\
\medskip
\textit{Supplemental Material}\\
\medskip
{Arnab Chaudhuri, Priya Mishra and Rukmani Mohanta }
\end{center}

\section{Derivation of the Lab-Frame Neutrino Flux}

For completeness, we summarize how the standard Hawking emission for a spinning (Kerr) primordial black hole (PBH) transforms from the rest frame to the observer’s lab frame when the PBH has bulk motion. This sets up the key expression used in our numerical evaluation.

\vspace{6pt}
\textbf{Rest-frame emission.}  
A Kerr PBH emits particles thermally, but the rotating horizon modifies both the energy spectrum and angular dependence. For a particle species of spin $s$, the differential emission rate per unit energy $E'$, time $t$, and solid angle $d\Omega'$ in the PBH rest frame is:
\begin{equation}
    \frac{d^3 N}{dE'\, dt\, d\Omega'} = 
    \sum_{\ell,m} |S_{\ell m}(\theta')|^2 \frac{\Gamma_{s,\ell m}(E',a_*)}{2\pi} 
    \frac{1}{\exp\!\big[(E' - m \Omega_H)/T_H\big] \pm 1},
\end{equation}
where $S_{\ell m}(\theta')$ are spin-weighted spheroidal harmonics and $\Gamma_{s,\ell m}$ are the greybody factors for each partial wave. The factor $(E' - m \Omega_H)$ reflects the horizon’s rotation:
\[
\Omega_H = \frac{a_* c^3}{2 G M (1 + \sqrt{1 - a_*^2})},
\quad
T_H = \frac{\hbar c^3}{8 \pi G M k_B} \frac{\sqrt{1 - a_*^2}}{1 + \sqrt{1 - a_*^2}}.
\]
The $m\,\Omega_H$ term favors co-rotating modes, leading to angular anisotropy in the rest frame.

\vspace{6pt}
\textbf{Lorentz boost to the lab frame.}  
If the PBH has a bulk velocity $\beta$ relative to the cosmic frame, the rest-frame emission must be Lorentz-transformed. For a boost along the $z$-axis:
\begin{align} 
    E &= \gamma E' (1 + \beta \cos\theta'),\\
    \cos\theta &= \frac{\cos\theta' + \beta}{1 + \beta \cos\theta'},\\
    d\Omega &= \frac{d\Omega'}{\gamma^2 (1 + \beta \cos\theta')^2},
\end{align}
where $\gamma = (1 - \beta^2)^{-1/2}$ is the Lorentz factor. This transformation accounts for both the Doppler shift in energy and the aberration of angles.

\vspace{6pt}
\textbf{Final lab-frame flux.}  
Combining the rest-frame emission spectrum with the Lorentz boost yields the full lab-frame differential neutrino flux:
\begin{align}
\frac{d^3 N}{dE\, dt\, d\Omega} &= 
\sum_{\ell,m} |S_{\ell m}(\theta')|^2\, 
\frac{\Gamma_{s,\ell m}(E',a_*)}{2\pi} 
\,\frac{1}{\exp\!\big[(E' - m \Omega_H)/T_H\big] + 1} \nonumber\\
&\times \frac{1}{\gamma^2 (1 + \beta \cos\theta')^2}\,
\delta\!\big(E - \gamma E'(1 + \beta \cos\theta')\big).
\end{align}

This expression encodes three key physical ingredients:

\begin{itemize}
    \item \textbf{1) Angular structure.}  
    The term $|S_{\ell m}(\theta')|^2$ represents the spin-weighted spheroidal harmonics that determine how the emission varies with polar angle in the PBH rest frame. For non-zero spin $a_*$, frame dragging favors modes with high $|m|$, enhancing emission in the equatorial plane.
    
    \item \textbf{2) Greybody factors and horizon rotation.}  
    The greybody factor $\Gamma_{s,\ell m}(E',a_*)$ accounts for the partial transmission probability through the black hole's curved spacetime. The factor $[E' - m \Omega_H]$ inside the exponent reflects how co-rotating modes are boosted in energy by the horizon’s rotation, distorting the thermal spectrum and favoring emission aligned with the PBH’s spin.
    
    \item \textbf{3) Doppler and aberration effects.}  
    The Lorentz boost modifies both energy and angle: the delta function enforces that only rest-frame neutrinos with energy $E'$ and angle $\theta'$ map to the lab-frame energy $E$ via the Doppler shift $E = \gamma E'(1 + \beta \cos\theta')$. The factor $1/[\gamma^2 (1 + \beta \cos\theta')^2]$ comes from the solid angle transformation (aberration) and compresses the emission into a narrower forward cone, enhancing the apparent flux for observers aligned with the PBH’s motion.
\end{itemize}

Physically, the combination of spin-induced angular anisotropy and bulk relativistic motion yields a lab-frame neutrino flux that is highly collimated and spectrally hardened relative to the isotropic, non-spinning case. This leads to a strong forward beaming of high-energy neutrinos, increasing the chance of detection by neutrino telescopes for observers located within the beaming cone.

In practice, we evaluate this by discretizing $E'$ and $\theta'$, then mapping to lab-frame $(E,\theta)$ to obtain the observable flux and angular distribution. The resulting spectrum shows clear signatures of both spin-induced anisotropy and relativistic boosting, as detailed in the main text.

\section{Numerical Implementation}

The numerical evaluation of the lab-frame neutrino flux requires accurate computation of the spin-weighted spheroidal harmonics \( S_{\ell m}(\theta') \) and the greybody factors \( \Gamma_{s,\ell m}(E,a_*) \), both of which depend sensitively on the black hole spin parameter \( a_* \) and the emitted particle’s energy.

\vspace{6pt}
\noindent
\textbf{Spin-weighted spheroidal harmonics:}  
We compute \( S_{\ell m}(\theta'; a_* \omega) \) by employing a spectral decomposition method, which expands the spheroidal harmonics in terms of spin-weighted spherical harmonics \( {}_s Y_{\ell' m}(\theta') \):
\begin{equation}
    S_{\ell m}(\theta'; a_* \omega) = \sum_{\ell' = |s|}^{\ell_{\max}} c_{\ell' \ell m}(a_* \omega) \, {}_s Y_{\ell' m}(\theta').
\end{equation}
The coefficients \( c_{\ell' \ell m} \) are obtained by solving the angular Teukolsky eigenvalue problem as a matrix eigenvalue problem. To ensure numerical convergence and accuracy, we truncate the series at \(\ell_{\max} = 6\), which is sufficient for the energy and spin ranges considered here. This truncation strikes a balance between computational efficiency and precision, as higher \(\ell\) modes contribute negligibly to the angular distribution for typical PBH parameters.

\vspace{6pt}
\noindent
\textbf{Greybody factors:}  
The greybody factors \( \Gamma_{s,\ell m}(E,a_*) \) represent the transmission probabilities for particles to escape the curved black hole spacetime and are crucial in shaping the emitted spectrum. For computational efficiency, we interpolate these factors from high-precision tabulated datasets available in the literature:
\begin{itemize}
    \item Page~\cite{Page:1976df} computed greybody factors for scalar, fermion, and vector fields over a wide range of black hole spins and rotation parameters, providing the foundational numerical results for Kerr black hole evaporation.
    \item MacGibbon and Webber~\cite{MacGibbon:1990zk} extended these results with improved numerical techniques and provided interpolation formulas spanning a broad energy range, facilitating efficient implementation in particle emission models.
\end{itemize}

Interpolation is performed on a finely spaced grid of energy \( E \) and spin \( a_* \), ensuring smooth, accurate evaluations during the flux calculation.

\vspace{6pt}
\noindent
\textbf{Numerical verification:}  
To validate the interpolation approach, we perform benchmark computations of \( \Gamma_{s,\ell m}(E,a_*) \) by directly solving the radial Teukolsky equation numerically for selected parameters. This involves integrating the radial wave equation with appropriate ingoing/outgoing boundary conditions and extracting transmission coefficients. These direct numerical results agree with the interpolated values within numerical uncertainties, providing confidence in our implementation.

\vspace{6pt}
\noindent
\textbf{Boost implementation and flux calculation:}  
The lab-frame neutrino flux is computed by discretizing the rest-frame energy \( E' \) and polar angle \( \theta' \) on sufficiently fine grids. The Lorentz transformation equations
\[
E = \gamma E'(1 + \beta \cos\theta'), \quad
\cos\theta = \frac{\cos\theta' + \beta}{1 + \beta \cos\theta'}
\]
are applied to map rest-frame variables to the lab frame. The Jacobian factor from the solid angle transformation is included to conserve particle number.

The delta function enforcing energy conservation in the boost is handled by binning \( E \) values and summing over contributing \( E' \) bins, weighted by the appropriate Jacobian factors. Angular integrals over \( \theta' \) use Gaussian quadrature to achieve high accuracy.

\vspace{6pt}
\noindent
This procedure yields the final lab-frame differential neutrino flux with full inclusion of spin-induced anisotropies, greybody filtering, and relativistic beaming due to PBH bulk motion.

The boost is implemented by mapping $(E',\theta')$ to lab-frame $(E,\theta)$ on a discrete grid. The delta function is handled by conserving flux: for each $E'$, the corresponding $E$ is binned with proper Jacobian factors.

\section{Effect of PBH Velocity on Neutrino Beaming Angle}

\begin{figure}[h]
\centering
\includegraphics[width=0.7\linewidth]{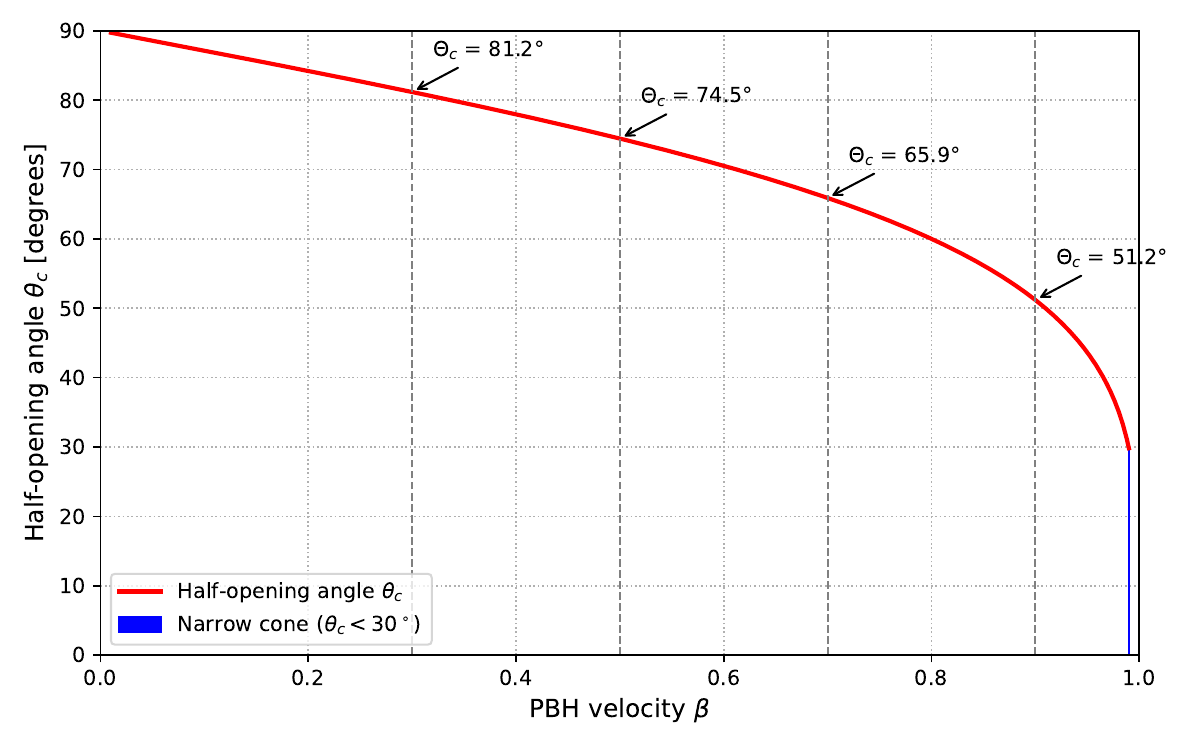}
\caption{Half-opening angle $\theta_c$ of the beaming cone vs.\ PBH velocity $\beta$. For $\beta \gtrsim 0.5$, the cone narrows to $\sim 40^\circ$.}
\label{fig:beaming_angle}
\end{figure}

Figure~\ref{fig:beaming_angle} illustrates in detail how the half-opening angle $\theta_c$ of the neutrino beaming cone depends on the bulk velocity of the PBH $\beta = v/c$.  
Here, $\theta_c$ is defined as the polar angle measured from the PBH’s direction of motion within which a fixed fraction (for example, 90\%) of the total neutrino flux is emitted.

In the PBH rest frame, Hawking emission is nearly isotropic in the absence of spin.  
However, when the PBH moves with velocity $\beta$ relative to the observer, the neutrino flux is transformed by a Lorentz boost along the direction of motion (taken to be the $z$-axis).  
This transformation modifies both the energy and angular distribution of the emitted neutrinos:
\begin{align}
E &= \gamma E' (1 + \beta \cos\theta'), \\
\cos\theta &= \frac{\cos\theta' + \beta}{1 + \beta \cos\theta'}, \\
d\Omega &= \frac{d\Omega'}{\gamma^2 (1 + \beta \cos\theta')^2},
\end{align}
where $E'$ and $\theta'$ are the neutrino energy and emission angle in the PBH rest frame, $\gamma = (1 - \beta^2)^{-1/2}$ is the Lorentz factor, and $d\Omega$ and $d\Omega'$ are the solid angles in the lab and rest frames, respectively.

The Doppler boost shifts neutrinos emitted in the forward direction ($\cos\theta' \approx 1$) to higher lab-frame energies, while aberration compresses their emission angles toward the direction of motion.  
As a result, the lab-frame angular distribution becomes increasingly peaked at small polar angles $\theta$ relative to the PBH's velocity vector.

At small velocities ($\beta \lesssim 0.2$), the effect is mild: the boosted emission remains nearly isotropic, and the half-opening angle $\theta_c$ enclosing 90\% of the flux stays close to $90^\circ$.  
However, as the PBH approaches moderate relativistic speeds ($\beta \gtrsim 0.5$), these relativistic effects become significant.  
The emission becomes strongly forward-collimated, and $\theta_c$ sharply decreases to around $40^\circ$ for $\beta=0.5$–$0.6$.

To quantify this effect, we numerically evaluate the lab-frame flux by integrating over the rest-frame angular distribution and applying the Lorentz transformation.  
The fraction of flux within a cone of half-opening angle $\theta_c$ is given by:
\begin{equation}
F(\theta_c) = \frac{\displaystyle \int_{0}^{\theta_c} \frac{d^2N}{dt\, d\Omega} \, 2\pi \sin\theta\, d\theta}{\displaystyle \int_{0}^{\pi} \frac{d^2N}{dt\, d\Omega} \, 2\pi \sin\theta\, d\theta}.
\end{equation}
By solving $F(\theta_c)=0.9$, we determine $\theta_c$ for each value of $\beta$.

This velocity-dependent collimation has a direct observational consequence: it increases the apparent neutrino flux for observers located inside the beaming cone, thereby enhancing the expected event rate in neutrino telescopes such as IceCube and KM3NeT.  
Even modest relativistic bulk velocities can thus significantly improve the detectability of PBH neutrino bursts by concentrating the emission into a narrower angular region.

Figure~\ref{fig:beaming_angle} quantitatively illustrates how PBH bulk motion leads to directional beaming of neutrino emission via Doppler boosting and aberration, complementing the spin-induced anisotropies discussed earlier and together predicting sharp, directional neutrino bursts from spinning, moving PBHs.

\section{Parameter Benchmarks}

Table~\ref{tab:benchmarks} summarizes the benchmark parameter values employed in our numerical evaluation of the lab-frame neutrino flux.
We select spin parameters up to $a_* = 0.9$, motivated by scenarios where PBHs retain significant angular momentum; bulk velocities up to $\beta=0.6$, corresponding to plausible peculiar motions from clustering or formation; and PBH masses in the range $10^9$–$10^{16}$ g, covering the window relevant for final evaporation today.
The energy and angular grid choices ensure numerical convergence of the boosted flux and angular beaming profiles.

\begin{table}[h]
\centering
\begin{tabular}{cccc}
\hline\hline
Parameter & Value(s) & Description \\ \hline
$a_*$ & 0.0, 0.5, 0.9 & Dimensionless spin \\
$\beta$ & 0.0, 0.3, 0.6 & Bulk velocity ($v/c$) \\
$M_{\mathrm{PBH}}$ & $10^{9} - 10^{16}~$g & PBH mass range \\
Energy grid & 0.1--500 GeV & Lab-frame energies \\
Angle grid & 100 points & $\cos\theta$ resolution \\
\hline\hline
\end{tabular}
\caption{Key benchmark parameters for spin, velocity, and numerical grids.}
\label{tab:benchmarks}
\end{table}

\end{document}